# Optical measurements of three-dimensional microscopic temperature distributions around gold nanorods excited by localized surface plasmon resonance

JunTaek Oh[1], Gu-Haeng Lee[2], Jinsung Rho[3], Seungwoo Shin[4,5], Bong Jae Lee[2,*],

Yoonkey Nam[3,*], and YongKeun Park[1,4,5,*]

[1]*Department of Physics, Korea Advanced Institute of Science and Technology (KAIST), Daejeon 34141, Republic of Korea*
[2]*Department of Mechanical Engineering, KAIST, Daejeon 34141, Republic of Korea*
[3]*Department of Bio and Brain Engineering, KAIST, Daejeon 34141, Republic of Korea*
[4]*KAIST Institute for Health Science and Technology, KAIST, Daejeon 34141, Republic of Korea*
[5]*TomoCube Inc., Daejeon 34051, Republic of Korea*
[6]*Current Affiliation: Samsung Electro-Mechanics, Suwon 16674, Republic of Korea*

The measurement and control of the temperature in microscopic systems, which are increasingly required in diverse applications, are fundamentally important. Yet, the measurement of the three-dimensional (3D) temperature distribution in microscopic systems has not been demonstrated. Here, we propose and experimentally demonstrate the measurement of the 3D temperature distribution by exploiting the temperature dependency of the refractive index (RI). Measurement of the RI distribution of water makes it possible to quantitatively obtain its 3D temperature distribution above a glass substrate coated with gold nanorods with sub-micrometer resolution, in a temperature range of 100°C and with a sensitivity of 2.88°C. The 3D temperature distributions that are obtained enable various thermodynamic properties including the maximum temperature, heat flux, and thermal conductivity to be extracted and analyzed quantitatively.

## I. INTRODUCTION

Three-dimensional (3D) temperature distribution, one of the essential physical quantities for indicating the state of a system, provides invaluable insights of diverse thermal-induced applications in micrometer-sized systems, including photothermal therapy [1], modulation of neural activity [2,3], drug delivery [4] and microfluidics [5]. Previously, to identify the temperature change of the systems, a number of approaches have been developed, such as infrared thermometry [6-8], fluorescent microscopy [9-12] and quantitative phase microscopy and interferometric microscopy [13-16]. However, previous tools only measured 2D temperature distribution [6] with labeling agencies [9-12] or solved heat diffusion equation to retrieve both 2D and 3D temperature distributions with exploiting assumptions about the physical properties of the system and regularization process with empirical parameters [13,14].

To overcome the limitations, we propose a new straightforward approach to measure the 3D temperature distribution by exploiting the temperature dependency of water refractive index (RI) [17]. Optical diffraction tomography (ODT) was employed to reconstruct the 3D RI distribution of heated water with multiple illuminations at various incident angles [18-20]. The temperature dependency of the RI of water allows converting the reconstructed 3D RI distribution into the 3D temperature distribution of water above a glass substrate coated with gold nanorods (GNRs), without solving the steady-state heat diffusion equation or employing complex assumptions. We experimentally demonstrated the feasibility of the proposed method with statistical analysis and the numerical simulation which solves the steady-state heat diffusion equation. We verified that the measured 3D temperature distributions provide diverse thermodynamic properties of the system, including the maximum temperature, heat flux, and thermal conductivity.

## II. PRINCIPLE

The main concept of three-dimensional temperature measurement is depicted in Figure 1. The measurement exploits the temperature dependency of the RI of water. In practice, this involves obtaining the 3D temperature distribution of hot water from the 3D distribution of the RI, which is reconstructed using the principle of ODT [18,19]. The temperature dependency of the RI of water can be expressed using an empirical equation,

$$\Delta T = \sum_i \beta_i \Delta n^i, \qquad (1)$$

where $\Delta T$ is the temperature change, $\Delta n$ is the RI change of water, and $\beta_i$ are empirical coefficients [17] (see Appendix A). The proposed concept is experimentally demonstrated by using GNRs and by generating heat in the water using localized surface plasmon resonance[21]. The GNRs are synthesized for the maximum absorption efficiency at the wavelength of the excitation source. Distilled water is loaded between two cover glasses, and the bottom cover glass is coated with GNRs to control the temperature distribution of water [22,23] [Figs. 1(b)-(d)].

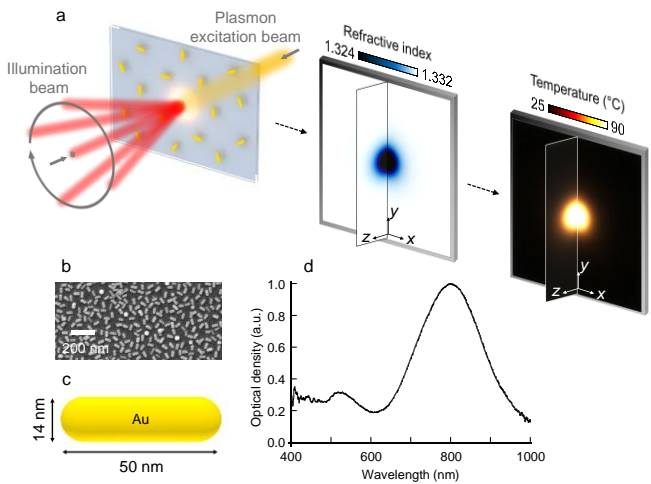

FIG.1. (a) A plasmon excitation beam (the yellow line) is focused on the bottom cover glass coated with GNRs to raise the temperature of the water. The holograms measured at various illumination angles (red lines) are used to reconstruct the 3D RI distribution of the heated, from which the 3D temperature distribution to be obtained. (b) SEM image of measured GNRs coated sample. (c) Specification of GNRs. 307 GNRs are statistically analyzed. The mean diameter, length and aspect ratio are 14.4 nm, 50.1 nm and 3.60, respectively. (d) The absorption spectrum of GNRs-coated cover glass.

## III. METHODS

### A. Experimental Setup

ODT, which is employed to measure the RI distribution of water in three dimensions [18,24-26], reconstructs the 3D RI distribution by measuring multiple 2D holograms at various illumination angles [24]. These holograms are acquired by using a Mach-Zehnder interferometer equipped with a digital micromirror device (DMD) [27]. A continuous wave (CW) infrared (IR) laser diode is utilized to excite the GNRs on the bottom glass cover and raise the temperature of the fluidic system (Figure 2a).

The optical fields diffracted by heated water at various illumination angles are retrieved from the obtained multiple holograms based on the Fourier transform method [28] (Figure 2b). The 3D RI distribution of hot water is constructed by mapping the retrieved optical fields onto 3D Fourier space based on the principle of ODT [18,29] (Figure 2c). Because of the limited numerical aperture of the objective lens, part of the information along the axial direction in the initially mapped data is missing[30]. In order to compensate for this limitation, we used an iterative regularisation method with two physical constraints: i) the reconstructed RI value of water cannot be higher than that of water at room temperature, and ii) the RI of the glass substrate remains constant because the changes in the RI of glass within the changed temperature range are negligible [17,31] (see Appendices B-C). Using these two constraints, the missing information in the Fourier space was filled (Figure 2c). Because the region of initially mapped data in the Fourier space was modified as well, the initial data was recovered, and an approximate solution was obtained [30]. We repeated this iterative regularisation process to obtain a converged 3D RI tomogram as illustrated in Figure 2d. As the last step, we reconstructed the 3D temperature distribution using eq 1 (Figure 2e).

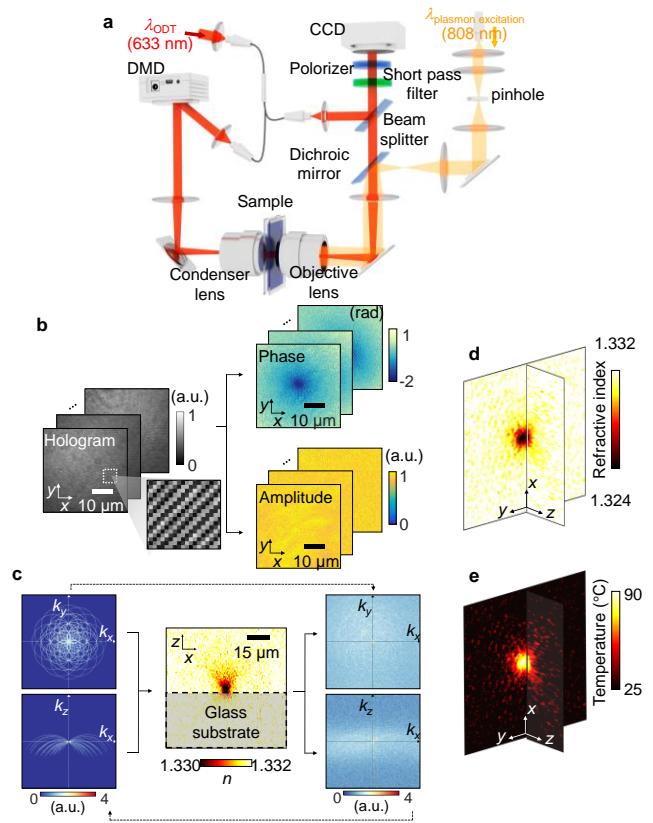

FIG.2. (a) ODT setup for the experiment. (b) The complex optical fields, which consist of the amplitude and phase, were retrieved from multiple holograms at various illumination angles. (c) The optical fields were projected onto 3D Fourier space. Missing information originating from the limited numerical aperture of the objective lens was filled in with an iterative regularisation algorithm using physical constraints. The 3D temperature distribution (e) was obtained from the measured 3D RI distribution (d) using eq 1.

The 3D RI distribution of a sample was reconstructed by employing the principle of optical diffraction tomography (ODT) [18,24]. A commercial ODT setup (HT-1H, Tomocube

Inc., Republic of Korea) was customized for the experiment. A HeNe laser (HNL150R, λ = 633 nm, 15 mW, Thorlabs Inc., USA) was used as a light source. The beam was split into two branches, the reference and sample arms, respectively. A digital micromirror device (DMD) (DLP LightCrafter 6500, Texas Instruments Inc., USA) was exploited to control the incident angles of the sample beam, and 45 different illumination angles were scanned at the frame rate of 15 Hz, which implies that the total acquisition time of one RI tomogram is 3 seconds. The modulated sample beams were demagnified by a water immersion objective lens [numerical aperture (NA) = 1.1 LUMPLN, 60×, Olympus Inc., Japan] and a tube lens ($f$ = 250 mm). Another high NA water immersion objective lens (NA = 1.2 UPLSAPO, 60×, Olympus Inc., Japan) and a second tube lens ($f$ = 180 mm) were used to collect the complex optical fields diffracted from the sample. The diffracted fields interfered with the reference beam on a CMOS image sensor (FL3-U3-13Y3M-C, FLIR Systems Inc., USA) and the image sensor records spatially modulated multiple holograms. An infrared laser diode (LD) was used as an excitation source (M9-808-015, λ = 808 nm, 150 mW, Thorlabs Inc., USA). The LD was mounted on a temperature controlled mount (TCLDM9, Thorlabs Inc., USA) and collimated using a laser collimation lens (F230FC-780, $f$ = 4.51 mm, Thorlabs Inc., USA). The LD beam was spatially filtered using a 30 μm pinhole (P30S, Thorlabs Inc., USA) and demagnified by a set of convex lenses ($f$ = 250 mm, 20 mm, and 100 mm) and a second tube lens ($f$ = 180mm).

### B. Control of illumination angles using a DMD

A DMD, which consists of millions of micromirrors, was used to control the illumination angles of the sample beam by displaying a binary grating pattern, known as a Lee hologram [32,33]. The first-order diffracted beam from the binary grating was spatially filtered and used as the illumination beam. Because of the binary image on a DMD, unwanted diffracted beams are inevitable and deteriorate the quality of the measurements. To overcome the limitation, we applied a time-multiplexing illumination technique [34], which displays a sequence of discrete binary patterns to express one continuous image. In particular, to display time-averaged sinusoidal patterns in this experiment, we consecutively displayed four decomposed binary images for one hologram. The obtained four holograms were numerically summed by considering the temporal weights of each frame to describe an approximate sinusoidal DMD pattern. As a result, we obtained 45 holograms and confirmed that the unwanted diffraction peaks in Fourier space were effectively reduced. The standard deviation of the RI distribution in the background region of the reconstructed tomograms decreases to the level of $10^{-4}$. This suppression is five times larger compared to the tomogram measured with binary patterns on DMD [35,36].

### C. Synthesis of gold nanorods

GNRs were synthesized by the seed-mediated method [37]. The seed solution was prepared by mixing 2.5 mL of 0.2 M cetyltrimethylammonium bromide (CTAB, Sigma), 2.5 mL of 0.5 mM HAuCl4 (Sigma), and 300 μL of ice-cold 0.01 M NaBH4 (Sigma) in an ultrasonication bath at room temperature. This seed solution was aged for 2 hours. The seeds were left in the growth solution at room temperature for 30 minutes to grow into rod-shaped structures. The growth solution consisted of a mixture of 5 mL of 0.2 M CTAB, 5 mL of 1 mM HAuCl4, 250 μL of 4 mM AgNO3 (Sigma), 70 μL of 78.84 mM ascorbic acid (Sigma) and 12 μL of seed solution. The GNRs were concentrated by centrifuge at 10,000 RPM (10200 RCF) and resuspended in ultrapure water to remove the surfactant. The GNRs were coated with polyethylene glycol (mPEG(5k)-SH, Nanocs) such that the GNR reached a ratio 3 optical density (O.D.) and 3 mg/mL of an aqueous solution of PEG for 12 hours at room temperature. Free PEG was removed using a dialysis kit (Thermo Scientific) for 2 days. The zeta potential of PEG-coated GNR was measured by a Zetasizer Nano ZS (Malvern).

Glass substrates were cleaned by successive ultrasonication in acetone, isopropyl alcohol, and deionized water, respectively, for 5 min each. Layer-by-layer coating was performed on the substrate to create a positively charged surface. Then, 10 mg/mL of poly(sodium 4-stirensulfonate) (PSS, MW~15,000, Aldrich) and poly(allylamine hydrochloride) (PAH, MW~200,000, Aldrich) were prepared in 10 mM NaCl solution. The substrates were treated with three cycles of PSS and PAH solution for 5 min each, ending with PAH. Finally, 0.2 mL/cm2 of 1 O.D. negatively charged GNR with a zeta potential -37.5 mV solution was loaded on the substrates for 12 hours.

The optical property of GNR-coated glass was measured by absorbance spectroscopy (USB4000-VIS-NIR-ES, Ocean Optics). The coverage of the GNRs are calculated as 30.6%. The measurement conditions were controlled to be same as for the experiment. The central absorbance peak was detected at approximately 808 nm, which was the wavelength of the stimulation laser. The average absorbance of the samples is 0.127 O.D, which implies that the extinction ratio of the sample is 25.4% for 808 nm wavelength. For the used GNRs with 14.4 nm diameter and 50.1 nm length, the scattering is significantly lower than the absorption for 808 nm wavelength; the portion of scattering is 10% with respect to the absorption [38,39] (Fig. 1d).

### D. Numerical Simulation

Laser-induced temperature distributions were calculated using a finite element method (FEM), for which we used commercial software (ANSYS). Construction of the simulation model for the actual experiment required us to make two assumptions. First, the convection effect in water was ignored, and only the conduction of water was considered. The beam size was less than 10 μm, and the gap between the glass slides was as small as 40 μm. The corresponding Rayleigh number was estimated to be $10^{-4}$, which confirms that the dominant heat transfer mode is indeed heat diffusion rather than convection [40] (See Append D). Second, thermophysical properties of the GNR-coated glass was assumed to be the same as those of glass because the GNR occupied a much smaller volume than the glass. The boundary conditions of the analysis were implemented at an ambient temperature of 25°C to determine the temperature distribution of water with a depth of 40 μm. The total size of the water and glass model was 200 μm × 200 μm × 200 μm. A laser with an 8.5 μm diameter was modeled as the heat source. Since the power of the laser can be described by a Gaussian function, the Gaussian heat source was divided into ten discrete regions. The power of the discretized area was calculated and used as the input heat source.

## IV. RESULTS

### A. Experimental demonstration

The experimental results are shown in Figure 3. The *x-y* and *x-z* cross-section images of the reconstructed 3D RI and temperature distribution are presented in Figures 3a and b, respectively. The temperature sensitivity, which is the minimum distinguishable temperature change, was investigated by analyzing the RI values in the background volume statistically. The distribution of the reconstructed RI values was fit to the Gaussian, and the full-width half-maximum was retrieved as $3.41 \times 10^{-4}$, which indicates that the temperature sensitivity of our setup is about 2.88°C (Figure 3d). The experiment was carried out under specific experimental conditions: the excitation laser was focused on the layer consisting of the GNRs. The intensity distribution of the excitation beam was presumed to be Gaussian. The Gaussian waist and total power of the excitation beam absorbed by the GNRs were estimated as 8.5 μm and 0.67 mW, respectively.

Two groups of measurements were acquired to demonstrate the system in dynamic equilibrium and the reproducibility of the proposed method, respectively. The first group of measurements was obtained to validate that the heated water is in the steady state. At the same measurement position with the excitation laser on, 30 temperature distributions were repeatedly measured at the interval of 3 seconds.

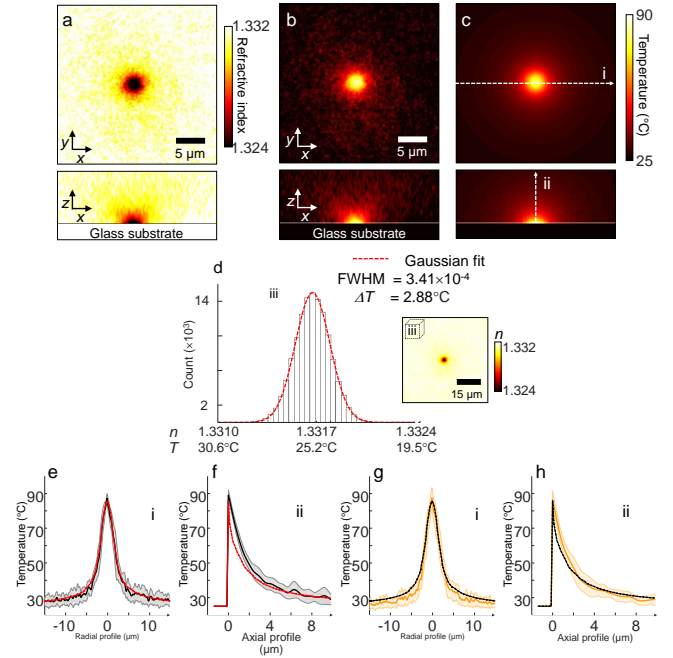

FIG.3. Experimentally measured 3D distribution of RI and temperature. (a) *x-y* and *x-z* cross-sectional images of 3D RI tomogram. (b) 3D temperature distribution obtained from 3D RI distribution. (c) Result of the numerical simulation solving the steady-state heat diffusion equation with assuming the same experimental conditions. (d) Distribution of RI values in background volume. (e,f) Temperature profiles along the radial and axial directions as white arrows (i) and (ii) as shown in Figure 3c. Thirty temperature distributions were measured at the same position for 3 minutes with the excitation laser turned on. The black line and areas represent the mean value and standard deviation, respectively. The red dashed line is the simulation result. (g,h) Temperature profiles extracted with the identical process used for (e) and (f). The yellow line and areas represent the mean value and standard deviation, respectively. The black dashed line represents the calculated values.

Figures 3e and f show the results of the measurements, extracted along the radial and axial directions indicated as white arrows (i) and (ii) in Figure 3c, respectively. The repetitive measurements were used to calculate the standard deviation (SD) of the maximum temperature as 2.92°C. This is comparable to the temperature sensitivity of the proposed method but slightly greater, which indicates that the system is in a quasi-steady state. The GNRs that are heated using the CW excitation source consistently supply energy to the water; thus, the temperature distribution reaches a steady state in a short time and the distributions measured over time remain constant [41]. The temporal deviation of the maximum temperature from the external noise of the system including fluctuations of the interferometry and excitation source.

The reproducibility of the proposed technique was verified by repeatedly measured 30 temperature distributions at 30 different measurement locations (Figs. 3g and h). The process that was used to extract the radial and axial profiles was identical to that used for the first group. The SD of the maximum temperature in this experiment was observed as 4.72°C. The fact that the SD is higher than for the first group implies the presence of another fluctuation in addition to the background noise of our setup. Because of the finite variance in the size and density of the GNRs coated on the bottom glass, the measured temperature distributions were modified depending on the position at which each measurement was conducted.

The temperature profiles of both the first and second groups deviated slightly from the theoretically computed values. For the axial profiles (Figs. 3f and h), the deviation between the experimental and theoretical values originates from the missing information along the optical axis. That is, the lack of axial information causes the reconstructed 3D RI distributions to have an elongated shape along the axial direction [42]. Therefore, the resulting 3D temperature distributions have slowly decreasing temperature profiles along the axial direction as illustrated in Figs. 3f and h.

A more in-depth analysis of the obtained 3D temperature distributions was carried out by measuring 30 temperature distributions for 8 different excitation powers. The maximum temperatures, and the distance from the origin to the isothermal positions at which the temperature increase is 25°C along the radial and axial directions were analyzed. The determination of the maximum temperature is in good agreement with the theory whereas both of the results of the isothermal distance deviate noticeably from the simulated results. The origin of these errors can be explained as discussed in Figs. 3e-h, considering that a greater amount of heat is transferred in the radial direction at the bottom of the water and taking into account the lack of axial information because the deviations have the same characteristics.

## B. Thermodynamic Analysis

We carried out a thermodynamic analysis including the 3D distribution of the heat flux density vectors and thermal conductivity of water (Fig. 4). These thermodynamic properties were calculated from the obtained 3D temperature distributions. The heat flux vector, which is defined as the amount of heat flow in unit time and unit area is expressed as,

$$\vec{q} = -\kappa \nabla T, \quad (2)$$

where $\vec{q}$ is the heat flux vector, $\kappa$ is the thermal conductivity of the medium and $\nabla T$ is the temperature gradient. Based on the obtained 3D temperature distribution, the heat flux vectors were visualized in 3D space to describe the way in which the heat is transferred to the surrounding medium as illustrated in Figs. 4d-g. We averaged 30 temperature distributions at each excitation power and assumed the thermal conductivity of water is constant [43] (0.6 W/m/K at room temperature) to reduce the noise and simplify the calculation, respectively.

On the other hand, the thermal conductivity can be estimated from the measured 3D temperature distributions (Fig. 4h). The integral of the heat flux vectors over an arbitrary closed surface which contains a heat source equals the power of the heat source delivered to the medium.

$$-\kappa \int_A \nabla T \cdot d\vec{a} = P, \quad (3)$$

where $P$ is the power delivered to the medium, $A$ is the arbitrary closed surface which contains the heat source. Because the integral of the temperature gradients can be calculated from the 3D temperature distribution, and the power delivered to the medium is obtained by multiplying the absorbance to the excitation laser power, one of the important thermodynamic properties, i.e., the thermal conductivity of water, can be estimated from the measured 3D temperature distributions employing eq 3. To simplify the calculation, we assumed an arbitrary cylindrical surface (see Appendix B). We calculated eight thermal conductivity values for each excitation power (Fig. 4i), which also shows the result of the calculations. These results show that the obtained values are comparable to the known thermal conductivity of water. The result implies that the assumption of the constant thermal conductivity of water, which was used to obtain the 3D distributions of the heat flux vectors, is appropriate as well.

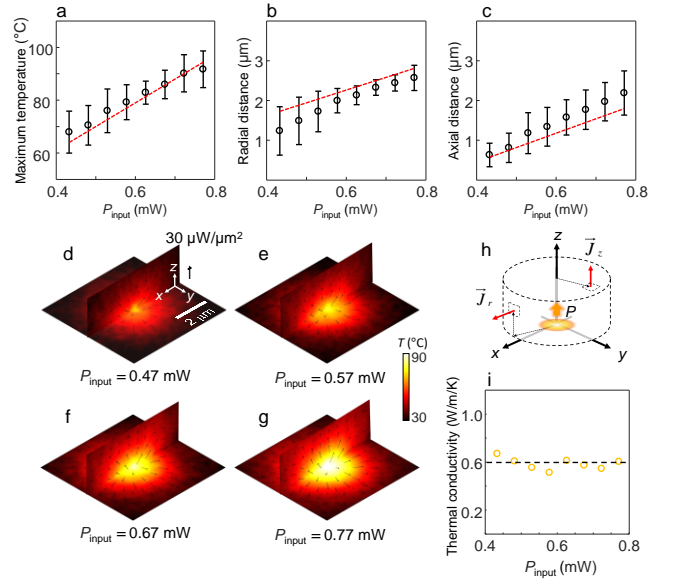

FIG.4. Quantitative analysis of various thermodynamic properties of 3D temperature distribution. The maximum

temperature (a) distances from the origin to the isothermal position at which the temperature increase is 25°C along the radial (b) and axial (c) directions were statistically analyzed. The thick black lines represent the experimental results. The red dashed lines show the result of the numerical simulation. (d-g) 3D distributions of the heat flux were obtained by taking the gradient of the 3D temperature distributions and employing the thermal conductivity of water. (h), Schematic of the calculation for the thermal conductivity of water. The total power delivered to both water and glass substrate, which is obtained by multiplying the absorbance of GNR sample, can be calculated by integrating the heat flux for an arbitrary closed surface *A*, which contains the heat source. The problem was simplified by specifying a cylindrical surface. (i) The result of the calculation is in good agreement with the known thermal conductivity values of water (the black dashed line, 0.6 W/m/K).

## V. DISCUSSION

In summary, we developed a new optical approach that enables the 3D temperature distribution of a microscopic water layer to be determined. The proposed method makes it possible to successfully visualize the 3D temperature distributions of hot water without any labeling agents. Our approach is straightforward and intuitive because the 3D temperature distributions are were reconstructed based on the measured axial information from multiple holograms at different illumination angles without solving the heat diffusion equations which requires assumptions and empirical regularization parameters. The only information required to measure the 3D temperature was the temperature dependency of the RI of water. Although we have demonstrated for the case of water as media, this approach is general and can be applied to other optically transparent media such as oil [44].

The measured 3D temperature distributions describe the comprehensive thermodynamic situation without considering any external variables. We suggested that the proposed method can be applied to identify thermodynamic properties such as the 3D distribution of the heat flux vectors and the thermal conductivity of an unknown fluidic system. On the other hand, improved temperature sensitivity can be achieved by employing a system with less noise, a common-path interferometric scheme [35]. The dynamic behavior of the system can be visualized by employing a faster acquisition and reconstruction configuration [45]. We envision that our method would facilitate an understanding of the various thermal phenomena in a microscopic fluidic system and foresee its use in various biomedical applications in the future.

## ACKNOWLEDGMENTS

Mr. Shin and Prof. Park have financial interests in Tomocube Inc., a company that commercializes ODT instruments and is one of the sponsors of the work. This work was supported by KAIST, BK21+ program, Tomocube, and National Research Foundation of Korea (2015R1A3A2066550, 2017M3C1A3013923, 2014K1A3A1A09063027, 2015R1A2A1A10055060, 2015R1A2A1A09003605). This research was also partly supported by KAIST Institute for the NanoCentury.

## APPENDIX A: Temperature dependency of RI change of water and glass substrate.

We referred to Schiebener P et al. [17] and Toyoda, T., and M. Yabe [31]. RI of water from 20°C to 90°C at 0.1 MPa (1 atm) and 632.8 nm was used (Table. 7 of Schiebener P et al.). The 3rd order polynomial fit curve (Figure 5) was estimated from the obtained RI values of water at each temperature and used to convert the 3D RI distribution into 3D temperature distribution.

At the interface between the glass substrate and water, the temperature profiles along the axial direction assumed to be continuous [41] so that we can compare the maximum RI change of both water and glass substrate of the same temperature increase. When the temperature of water increases from 25°C to 100°C, the refractive index (RI) change is approximately $|\Delta n_{water}| \sim 1.5 \times 10^{-2}$, while RI change of fused silica is estimated to be $|\Delta n_{glass}| \sim 9.3 \times 10^{-4}$. As a consequence, we noticed that the maximum RI change of glass substrate is less than 6% of RI change of water, which supports the assumption we used in the experiment: the RI change of glass substrate is negligible.

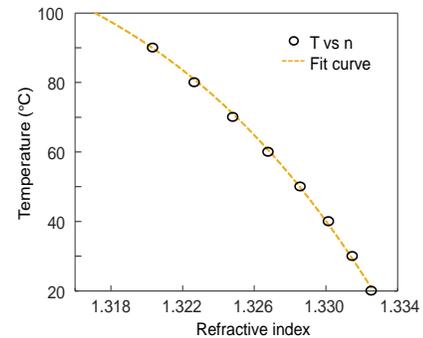

FIG.5. Quantitative analysis of various thermodynamic properties of 3D temperature distribution. Thirty temperature distributions were measured for each of eight different values of the excitation power

## APPENDIX B: Calculation of thermal conductivity.

Heat flux vector describes heat transfer in a medium and defined as the gradient of temperature distributions. A divergence of heat flux, on the other hand, indicates a power density of heat source. Because GNRs, two-dimensional heat source are deposited at the interface of water and glass substrate, the heat generated from the source must be transferred to both sides, and we noticed that the ratio of the total heat delivered to each material is determined by the ratio of the thermal conductivity of each medium [46,47].

$$P_{water} : P_{glass} = \kappa_{water} : \kappa_{glass}, \quad (4)$$

where $P_{water}$ is the power transferred to the water, $P_{glass}$ is the power delivered to the glass substrate, $\kappa_{glass}$ is the thermal conductivity of glass and $\kappa_{water}$ is the thermal conductivity of water that we are trying to calculate. With assuming that we do not have *a priori* information, the thermal conductivity of water, we defined a new experimental parameter $P_{exp}$ with omitting the unknown thermal conductivity in the equation (3).

$$P_{exp} = -\int_A \nabla T \cdot d\vec{a}. \quad (5)$$

where $A$ is now the surface region over the water, excluding the area of the glass substrate. From the known absorption efficiency of the heat source, we determined an initial input power $P_{input}$ which is absorbed and emitted by GNRs and substituted $P_{glass}$ to $P_{water}$ in Eq. 4. Combining Eq. 4 and 5, finally, we can estimate the unknown thermal conductivity value of water.

$$\kappa_{water} = \frac{P_{input}}{P_{exp}} - \kappa_{glass}. \quad (6)$$

We used 1.4 W/m/K, as the thermal conductivity of the glass substrate [31]. To simplify the calculation, we integrated heat flux vectors for an arbitrary cylindrical surface over the region of water.

## APPENDIX C: Contribution of heat conduction and convection

Rayleigh number (Ra) of the fluidic system implies the contribution of the convection to the heat transfer. Ra is defined as

$$Ra = \frac{VL}{\alpha}, \quad (7)$$

where $V$ is the velocity of the flow, $L$ is the size of the heat source and $\alpha$ is the diffusivity of the water [40]. In the practical experiment, V = $10^{-6}$ m/s, L = 8.5 μm and α = 1.4×$10^{-7}$ m$^2$/s. The calculated $Ra$ in the experiment is the order of $10^{-4}$ and it implies that the contribution of the convection to the temperature distribution is negligible.

## APPENDIX D: The timescale to the steady-state

In the experiment, the 3D temperature distribution of water reaches the steady-state in a quite short time according to the relation

$$\tau = \frac{L^2}{\alpha}, \quad (8)$$

where $L$ is the dimension of the heat source and $\alpha$ is the diffusivity of the water [40]. The intensity distribution of the heat source is 2D Gaussian of 8.5 μm waists, and the diffusivity of water is 1.4×10-7 m$^2$/s, which implies that $\tau$ is the order of $10^{-4}$ and the system reaches the steady-state in the time scale of microseconds. After heating, the temperature distribution remains constant because the continuous wave laser diode was employed in the practical experiment.

## APPENDIX E: The portion of absorption and scattering of GNRs

We simulated the boundary element method to solve the scattering and absorption problem of GNRs of 14.4 nm diameter and 50.1 nm length for 808 nm wave. The result indicates that the scattering efficiency of GNR is 10% of its absorption efficiency; thus, we can neglect the effect of scattering (Fig. 6).

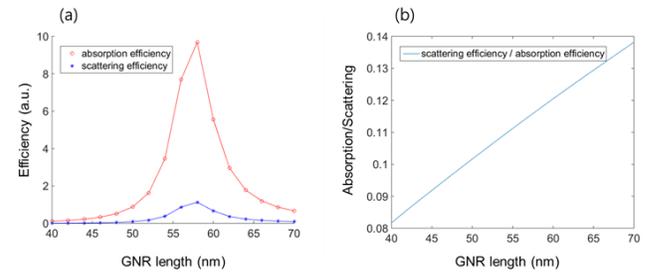

FIG.6. Quantitative analysis of various thermodynamic properties of 3D temperature distribution. Thirty temperature distributions were measured for each of eight different values of the excitation power